\renewcommand{\baselinestretch}{0.935}

\documentclass[sigconf]{acmart}
\usepackage[utf8]{inputenc}
\usepackage{enumitem}
\usepackage{multirow}
\usepackage{makecell}
\usepackage{balance}
\usepackage{subcaption}
\usepackage{tablefootnote}
\usepackage{dsfont}
\def\baselinestretch{0.935}
\usepackage[norelsize,linesnumbered,ruled,vlined,boxed,commentsnumbered]{algorithm2e}

\usepackage{graphicx} 
\usepackage{amsmath} 
\usepackage{booktabs}
\usepackage{enumitem}
\usepackage{multirow}
\usepackage{makecell}
\usepackage{balance} 
\usepackage{pifont}
\usepackage{bbding}
\usepackage{tikz}
\usepackage{multirow}
\usepackage{multicol}
\usepackage{xcolor}
\usepackage{tabularx}
\usepackage{subcaption}
\usepackage[all=normal,paragraphs=tight,floats=tight]{savetrees}

\setlist[itemize]{noitemsep, topsep=0pt, leftmargin=*}

\usetikzlibrary{arrows.meta, positioning}

\SetKw{KwOr}{or}
\SetKw{KwAnd}{and}

\title{Escaping Flatland: A Placement Flow for Enabling 3D FPGAs}

\copyrightyear{2026}
\acmYear{2026}
\setcopyright{cc}
\setcctype{by}
\acmConference[DAC '26]{63rd ACM/IEEE Design Automation Conference}{July 26--29, 2026}{Long Beach, CA, USA}
\acmBooktitle{63rd ACM/IEEE Design Automation Conference (DAC '26), July 26--29, 2026, Long Beach, CA, USA}
\acmDOI{10.1145/3770743.3804424}
\acmISBN{979-8-4007-2254-7/2026/07}

\author{Cong Hao}
\affiliation{%
\small
  \institution{Georgia Tech}
  \city{Atlanta, GA}
  \country{USA}
  }
\email{callie.hao@gatech.edu}

\author{Andrew B. Kahng}
\affiliation{%
  \small
  \institution{UC San Diego}
  \city{San Diego, CA}
  \country{USA}
  }
\email{abk@ucsd.edu}

\author{Bodhisatta Pramanik}
\affiliation{%
  \small
  \institution{UC San Diego}
  \city{San Diego, CA}
  \country{USA}
}
\email{bopramanik@ucsd.edu}

\author{Ismael Youssef\,}
\affiliation{%
\small
  \institution{Georgia Tech}
  \city{Atlanta, GA}
  \country{USA}
  }
  \authornote{Lead author.}
\email{ismael.youssef@gatech.edu}

\date{September 2025}

\begin{document}

\begin{abstract}
3D field-programmable gate arrays (FPGAs) promise higher performance through 
vertical integration. However, existing placement tools, largely inherited from 2D frameworks, fail to capture the unique delay characteristics and optimization dynamics of 3D 
fabrics. \textcolor{black}{We introduce} a 3D FPGA placement flow that integrates partitioning-based 
initialization, adaptive cost scheduling, refined delay estimation, and a simulated annealing move set –- all 
targeted at 3D FPGA architecture. Together, these enhancements improve timing estimates and the exploration 
of {\em layer} assignments during placement. Compared to Verilog-To-Routing (VTR), our experiments show geometric-mean (max) critical-path delay 
reductions of $\sim$3\% ($\sim$7\%), $\sim$2\% ($\sim$4\%), $\sim$3\% ($\sim$8\%), and $\sim$6\% ($\sim$18\%) 
for four 3D architectures: {\em 3D CB, 3D CB-O, 3D CB-I, and 3D SB}, respectively. We also achieve geometric-mean (max) 
routed wirelength reductions of $\sim$1\% ($\sim$3\%), $\sim$2\% ($\sim$8\%), $<1$\% ($\sim$5\%), and $\sim$5\% 
($\sim$10\%), respectively. Our work will be permissively open-sourced on GitHub.

\end{abstract}

\maketitle

\section{Introduction}

\begin{figure}[bt]
    \centering
    \vspace{2pt}
    \includegraphics[width=\columnwidth]{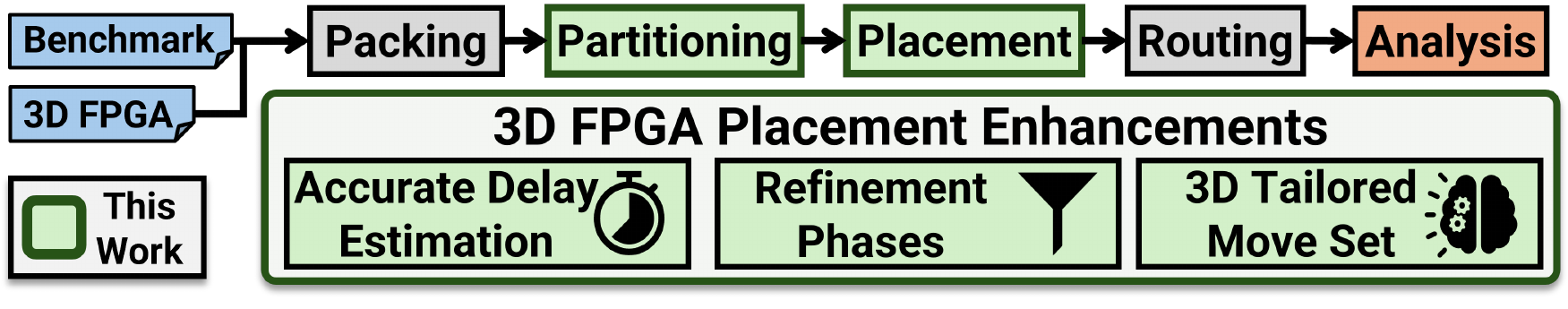}
    \vspace{-20pt}
    \caption{Overview of our 3D FPGA placement flow.}
    \label{fig:flow}
    \vspace{-18pt}
\end{figure}

FPGAs are central to modern computing, and \textcolor{black}{enable} domain-specific acceleration in applications 
ranging from AI and cloud services to medical imaging and wireless communication~\cite{military, pathway, KongJX15}. 
As computational demands grow, improving FPGA efficiency—e.g., computation density, total compute capacity, and 
achievable clock frequency—has become increasingly important. 3D integration offers a compelling path forward. By 
vertically stacking logic layers, 3D FPGAs can reduce wirelength and delay, increase logic density, and alleviate 
congestion and I/O limitations~\cite{ReifFCD02, Guarini02, LinGLW07}. Recent advances in monolithic integration, 
Through Silicon Vias (TSVs), and hybrid bonding have made such architectures practically realizable~\cite{SamalNIBK16, ChengGP22, xilinx_tsmc}.
However, Computer-Aided Design (CAD) support for 3D FPGAs remains limited. Existing 3D placement approaches~\cite{ElgammalMSMKT25, Rahimi23, 
AbabeiMB05} do not accurately model inter-layer delay, and perform \textcolor{black}{only} limited inter-layer placement exploration, 
as emphasized by~\cite{YoussefHH25}. To address these gaps, this work introduces 
a 3D FPGA placement flow that explicitly models inter-layer delay, has a tailored move set for 3D placement, and optimizes 
a dedicated placement cost for 3D FPGAs. The key contributions of our work are \textcolor{black}{as follows.}

\textbf{Open-source 3D FPGA placement flow. }
We propose a 3D FPGA placement flow (Figure~\ref{fig:flow}) that generates
high-quality placement solutions across multiple 3D architectures. In particular, we test our flow on the {\em 3D connection box (CB), 3D CB-O,
3D CB-I}, and {\em 3D switch box (SB)} architectures defined using the open-source 3D FPGA architecture exploration framework, LaZagna~\cite{YoussefHH25}, as well as hybrid variants of these. Our flow
is integrated into the open-source FPGA implementation framework VTR~\cite{vtr_repo}. At its core,
the flow leverages the hypergraph partitioner {\em TritonPart}~\cite{BustanyGKKPW23, tritonpart} to perform
layer {\em assignments}, and augments VTR's placement algorithm with 3D-specific operators (move set) (Section~\ref{sec:methodology}). 
\textcolor{black}{Our implementation is available on GitHub under a permissive open-source license~\cite{our_flow}}.

\textbf{3D placement enhancements. }
We introduce several enhancements to VTR to improve 3D placement quality and enable broader 3D architecture exploration.
First, {\em TritonPart}'s layer assignment serves as the initial solution for VTR's simulated annealing (SA)-based placer. 
Second, we propose a delay model that accurately captures vertical interconnect delay, providing stronger 
timing guidance during 3D placement. Third, we add
refinement phases during 3D placement: in early iterations, wirelength and congestion are prioritized and inter-layer moves
are penalized to preserve layer structure, while in later iterations inter-layer penalties are relaxed and timing is
prioritized to optimize critical paths across layers. Finally, we extend VTR's placement move operators to enable 
better layer exploration during 3D placement (Section~\ref{sec:methodology}).

\textbf{Evaluation and architecture exploration. }
We evaluate our proposed flow on the Koios design suite~\cite{AroraBDMMA23} across representative 3D FPGA architectures proposed by LaZagna~\cite{YoussefHH25}. Experimental results demonstrate that compared to the default VTR flow, our flow achieves
geometric-mean (max) critical-path \textcolor{black}{delay reductions of up to $\sim$6\% ($\sim$18\%)}. We also obtain geometric-mean (max) routed
wirelength reductions of up to $\sim$5\% ($\sim$10\%) (Section~\ref{sec:evaluation}). Routability studies show that with our enhancements, the flow can find lower minimum channel widths than
the default VTR implementation (Section~\ref{sec:methodology}). Moreover, our flow enables more \textcolor{black}{accurate and} effective architecture exploration than prior work
~\cite{vtr_repo, YoussefHH25}. Finally, our studies show that (i) our flow can be used for architectural and performance exploration for 3D FPGA designs, and (ii) \textcolor{black}{some 3D architectures previously considered underperforming relative to other 3D variants can in fact significantly outperform them when evaluated with our framework.}

The remainder of the paper is organized as follows.
Section~\ref{sec:related} reviews related work.
Section~\ref{sec:background_fpga_architectures} \textcolor{black}{presents 
background} on 3D FPGA architectures.
Section~\ref{sec:methodology} describes our proposed approach.
Section~\ref{sec:evaluation} presents experimental results.
Section~\ref{sec:conclusion} concludes the paper.

\section{Related work}
\label{sec:related}

FPGA placement has been extensively studied for 2D fabrics~\cite{RajarathnamAJIP22, ElgammalMB22, MahmoudiESMB23, MarquardtBR00}. 
In contrast, 3D FPGA placement remains relatively underexplored. Existing works can be broadly grouped into two categories.

\noindent \textbf{Layer assignment via partitioning.}
This strategy decomposes the placement problem into two stages: first, a layer-assignment phase that partitions 
the netlist into the target number of layers; and second, a conventional 2D placement flow \textcolor{black}{applied} independently 
within each layer~\cite{AbabeiMB05, ChtourouAMAM17, DanassisSS16, ZhaoAIY20, Rahimi23}. 
However, these works are evaluated primarily on small benchmark circuits and simpler {\em homogeneous} FPGA fabrics that 
lack hard blocks\footnote{A block in an FPGA is a placeable logic or memory resource (IO, CLB, DSP, BRAM). 
During placement, every block 
corresponds to an element in the {\em packed} netlist~\cite{ElgammalMSMKT25} that must be mapped to a 
physical site on the FPGA fabric.} 
such as DSPs and BRAMs. As a result, they are not directly applicable to modern {\em heterogeneous} 
FPGA architectures that combine {\em configurable logic blocks} (CLBs) with embedded memory and DSP resources\textcolor{black}{, and cannot be compared with this work}. Moreover, 
with the exception of~\cite{AbabeiMB05}, all previous implementations remain closed-source.

\noindent \textbf{Extending ``2D'' cost function and move space.}
A more recent line of research augments conventional 2D placement by introducing a $z$-coordinate into the
cost and delay models~\cite{BoutrosMMMB23}, effectively treating the ``third dimension'' as an additional spatial axis. 
\cite{BoutrosMMMB23} integrates such 3D support into the VTR framework,
extending its wirelength and timing models to capture inter-layer connectivity\footnote{In this work, we 
use the terms ({\em vertical, inter-layer, 3D}) connectivity interchangeably.} while retaining VTR's 2D annealing
schedule and move set.

In this work, we combine the benefits of both categories. We use a {\em partitioning-based initial layer assignment},
but do not restrict logic blocks to their assigned layers. Instead, blocks are allowed to move across layers 
under a dynamically adaptive cost function that preserves the initial layer structure (i.e., partitioning-based
layer assignment) in early placement iterations and enables timing-driven optimization across layers 
in later iterations (Section~\ref{sec:meth_var_cost}).

\section{3D FPGA Architectures}
\label{sec:background_fpga_architectures}

\begin{figure}[bt]
    \centering
    \vspace{-8pt}
    \includegraphics[width=1.0\columnwidth]{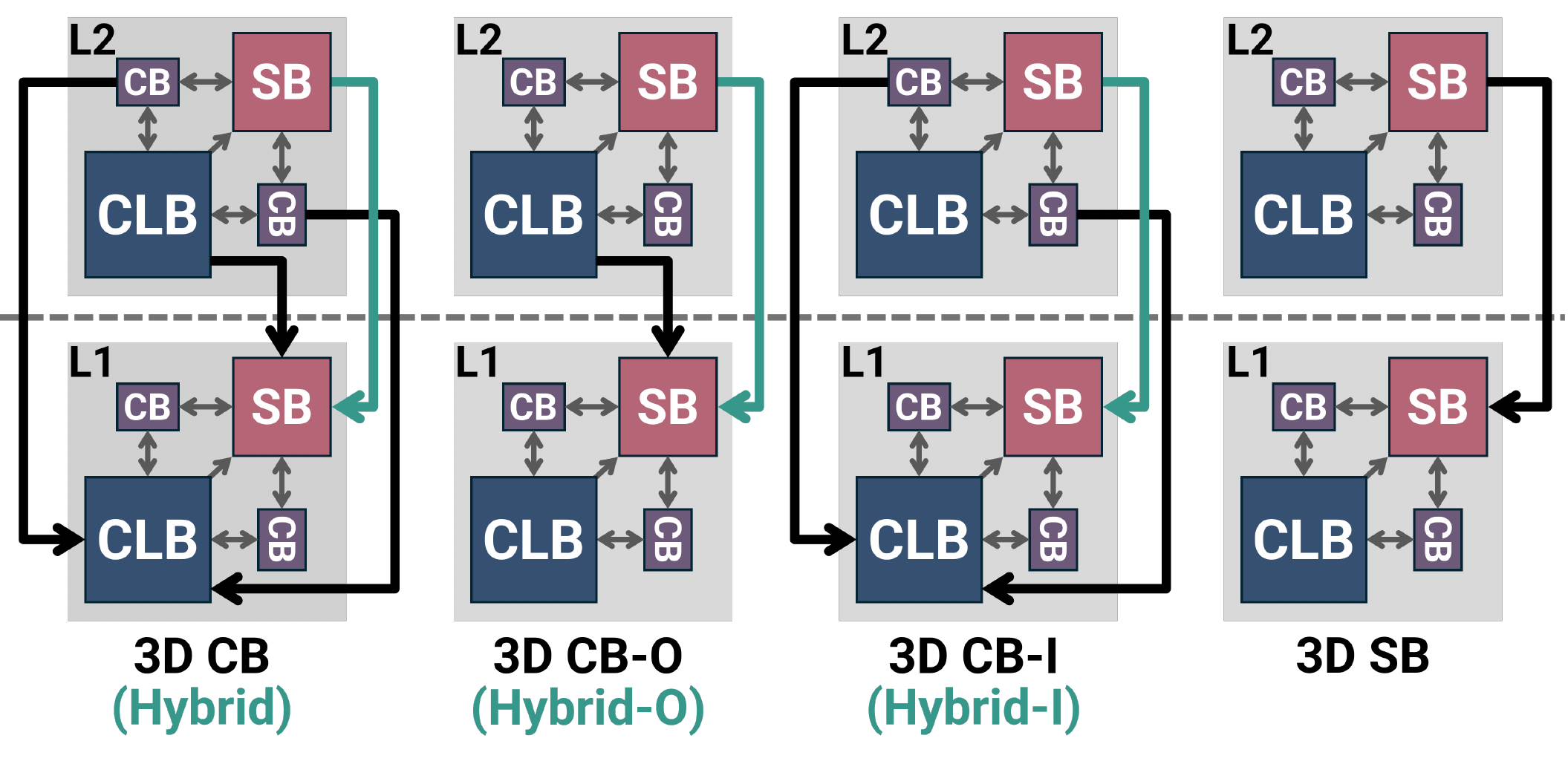}
    \vspace{-20pt}
    \caption{3D FPGA architectures based on connection types. 
    Connections are only shown from layer ``L2'' to layer ``L1''. 
    Green arrows show hybrid variations of each architecture.}
    \label{fig:archs}
    \vspace{-8pt}
\end{figure}

3D FPGAs can be architected in different ways depending on how logic blocks and routing resources are distributed
across layers and how vertical connections are inserted between them. 
Prior architectural studies~\cite{AmagasakiZIKS15, BoutrosMMMB23, LeRB09} have explored a range
of inter-layer connectivity patterns and shown that these choices strongly affect routability and timing.
Amagasaki et al.~\cite{AmagasakiZIKS15} propose architectures where logic blocks have dedicated inter-layer pins;
Boutros et al.~\cite{BoutrosMMMB23} examine architecture variants that restrict vertical connections to logic block outputs; and
Le et al.~\cite{LeRB09} investigate switch-block-based vertical routing schemes. 
More recently, Youssef et al.~\cite{YoussefHH25} perform a comprehensive sweep across different 3D connectivity models
and \textcolor{black}{evaluate} their benefits.
To capture this diversity, and to build on the architectures established by LaZagna~\cite{YoussefHH25},
we consider the same four representative architecture types (Figure~\ref{fig:archs}):
(i) \textit{3D CB:} inter-layer connections are made through {\em input and output pins} of logic blocks; 
(ii) \textit{3D CB-O:} inter-layer connections are made through {\em output pins} of logic blocks;
(iii) \textit{3D CB-I:} inter-layer connections are made through {\em input pins} of logic blocks; \textcolor{black}{and}
(iv) combinations of {\em 3D SB} connectivity with a {\em 3D CB}-based architecture 
({\em 3D CB-O, 3D CB-I, 3D CB})\textcolor{black}{, where} inter-layer connections are made through both {\em logic block pins and SB resources} 
(green arrows in Figure~\ref{fig:archs}).

The above configurations span the vertical-connectivity paradigms studied in previous work~\cite{YoussefHH25, BoutrosMMMB23}. More
architectures can be expressed as combinations of the above types. 
Our evaluation focuses on the above types with {\em two-layer} fabrics. 
This aligns with near-term 3D FPGA integration scenarios~\cite{Hoefflinger20} 
and captures the key placement challenges introduced by vertical connections.
\footnote{\textcolor{black}{Extending the flow to more than two layers is non-trivial and is left for future work. Unlike the two-layer case, deeper stacks introduce fundamental reachability constraints. For example, if only certain pins support vertical connections, nets between non-adjacent layers (e.g., layer 1 to layer 3) may be unroutable without intermediate resources. Addressing this would require changes to either or both of the routing architecture and placement formulation, which are beyond the scope of this work.}}

\section{Our approach}
\label{sec:methodology}

\begin{table}[t]
\centering
\scriptsize
\caption{Notations and terminologies.}
\vspace{-8pt}
\label{tab:notation}
\begin{tabularx}{\columnwidth}{lX}
\toprule
\textbf{Symbol} & \textbf{Definition} \\
\midrule
$b, \mathcal B$ & A block in the netlist and set of blocks in the netlist, respectively, where $b \in \mathcal B$. \\ 
$l_{\text{src}},\, l_{\text{dst}}$ & Source and destination layer indices for a move. \\
$\Delta x,\, \Delta y$ & \textcolor{black}{Absolute} Manhattan (horizontal) displacement between source and destination grid locations. \\ 
\textcolor{black}{$T, T_{\text{init}}, T_{\text{exit}}$} & \textcolor{black}{Annealing temperature variables: current, initial (Section 9.1 in~\cite{ElgammalMSMKT25}), and exit temperature (Section 3 in~\cite{BetzR97}), respectively.} \\ 
$\text{delay}$ & Lookup table entry storing precomputed interconnect delay for \textcolor{black}{a connection with} bounding-box from 
$(l_{\text{src}}, x, y)$ to $(l_{\text{dst}}, x+\Delta x, y+\Delta y)$. \\
$C_{\text{bb}}(b)$ & Bounding-box wirelength cost contribution of block $b$ (Equation 2.2 in \cite{Mohaghegh09}). \\
$C_{\text{bb}}$ & Total bounding-box wirelength cost of the placement (Equations 4.2, 4.3 in~\cite{Mohaghegh09}). \\
$C_{\text{timing}}(b)$ & Timing cost contribution of block $b$ (Equation~\ref{eq:timing_cost}). \\ 
$C_{\text{timing}}$ & Total timing cost of the placement (Equations 7, 8 in~\cite{MarquardtBR00}).\\
$C_{\text{total}}(b)$ & Total placement cost per block $b$ (Equation~\ref{eq:total_cost}). \\ 
$C_{\text{total}}$ & Aggregated total placement cost (SA cost function): $\sum_{b \in  \mathcal B}{C_{\text{total}}(b)}$. \\ 
$N_{\text{moves}}$ & Total number of moves per temperature step. \\ 
$\alpha$ & Move-acceptance rate, i.e., the fraction of attempted moves (out of $N_{\text{moves}}$) that \textcolor{black}{are} accepted. \\ 
$w$ & Arithmetic average of move-acceptance rates in \textcolor{black}{the}
last five temperature steps. \\
$k$ &  Number of layers in the 3D FPGA fabric. \\
$N_{\text{blk}}$ & Number of movable logic blocks in the netlist. \\
$\theta(w)$ & Our {\em dynamic} timing vs. wirelength weighting factor. \\ 
$\zeta(w)$ &  Our {\em dynamic} scaling factor that increases penalty for inter-layer timing cost. \\ 
$p_{\zeta}, p_{\theta}$ & Exponential scaling factors (Equations~\ref{eq:alpha_dynamic},~\ref{eq:tt_equation}). \\ 
$\theta_{\min},\, \theta_{\max}$ & Preset bounds (min, max) of $\theta(w)$. \\
$\zeta_{\text{min}},\, \zeta_{\text{max}}$ & Preset bounds (min, max) of $\zeta(w)$. \\
$w_{\text{min}}, w_{\text{max}}$ & Preset bounds (min, max) of $w$. \\
\bottomrule
\end{tabularx}
\vspace{-8pt}
\end{table}

We \textcolor{black}{now} describe our 3D placement flow (\textcolor{black}{see} notations and terminologies in
Table~\ref{tab:notation}). The overall flow is shown in Figure~\ref{fig:flow}, with pseudocode in
Algorithm~\ref{alg:3d_place}. To enable reliable evaluation of 3D FPGA architectures, the placement
engine must (i) accurately capture vertical delay, (ii) adapt its optimization operators to layered
fabrics, and (iii) \textcolor{black}{efficiently} explore the 3D search space. This section details the modifications
introduced into the VTR placement engine to meet these goals. Each enhancement contributes
incrementally to placement quality; together\textcolor{black}{,} they form a cohesive 3D-aware flow (see ablation
studies in Section~\ref{subsec:ablation}). VTR uses simulated annealing~\cite{KirkpatrickGV83} as its
core placement algorithm, and we implement all of our enhancements on top of the reference
implementation in~\cite{vtr_repo}.

\begin{algorithm}
\scriptsize
\caption{3D-Aware Simulated-Annealing Placement.}
\label{alg:3d_place}
\KwIn{$\mathcal{N}$: netlist, $k$: number of layers, $\mathcal{M}$: SA move set, 
      $T_{\text{init}}$: initial temperature, $T_{\text{exit}}$: exit temperature}
\KwOut{$\textit{place}_{\textit{best}}$: best 3D placement found}
\tcp{\textcolor{blue}{Phase 1: Initialization and layer assignment}}
$\mathcal{N}_{\text{part}} \gets$ partition $\mathcal{N}$ into $k$ layers using {\em TritonPart} \\
$\textit{place}_{\textit{curr}} \gets$ construct an initial placement consistent with $\mathcal{N}_{\text{part}}$ \\
$T \gets T_{\text{init}}$ \tcp*{current temperature}
$\alpha \gets 1.0$ \tcp*{initial value of move-acceptance rate} 
$\theta(w) \gets \theta_{\text{min}}$ \tcp*{initial timing weight (Section ~\ref{sec:meth_var_cost})}
$\zeta(w) \gets \zeta_{\text{max}}$ \tcp*{initial 3D timing scale (Section ~\ref{sec:meth_var_cost})}
$w \gets 1.0$ \tcp*{initial value of average move-acceptance rate}
$N_{\text{blk}} \gets$ number of movable blocks in $\mathcal{N}$ \\
$N_{\text{moves}} \gets 0.5 \cdot N_{\text{blk}}^{4/3}$ \tcp*{moves per temperature step (VTR heuristic)} 
$\textit{place}_{\textit{best}} \gets \textit{place}_{\textit{curr}}$  \\
$C_{\text{BB}} \gets$ get bounding-box wirelength cost of $\textit{place}_{\textit{curr}}$ (see Table~\ref{tab:notation}) \\ 
$C_{\text{timing}} \gets$ get timing cost of $\textit{place}_{\textit{curr}}$ (Table~\ref{tab:notation}) \\
$C_{\text{best}} \gets$ create placement cost tuple $ \langle C_{\text{BB}}$, $C_{\text{timing}} \rangle$ \\ 
\tcp{\textcolor{blue}{Phase 2: Simulated-annealing-based 3D placement}}
\While{$T > T_{\text{exit}}$}{
  \For{$i \gets 1$ \KwTo $N_{\text{moves}}$}{
    $m \gets$ sample a move from $\mathcal{M}$ using the 3D move-selection policy (Section ~\ref{sec:meth_rl}) \\

    $\Delta C_{\text{total}} \gets$ compute change in $C_{\text{total}}$ (see Table~\ref{tab:notation}, Equation~\ref{eq:total_cost}) \\
    \If{$\Delta C_{\text{total}} < 0$ \KwOr $e^{-\Delta C_{\text{total}} / T} > \text{rand}(0,1)$}{
      $\textit{place}_{\textit{best}} \gets$ apply move $m$ to the current placement \\
      $C_{\text{BB}} \gets$ get bounding-box wirelength cost of $\textit{place}_{\textit{curr}}$ (Table~\ref{tab:notation}) \\ 
      $C_{\text{timing}} \gets$ get timing cost of $\text{place}_{\text{curr}}$ (see Table~\ref{tab:notation}) \\
      $C_{\text{curr}} \gets$ create placement cost tuple $  \langle C_{BB}, C_{timing} \rangle$ \\ 
      \If{$C_{\text{curr}}.{C_{BB}}$ < $C_{\text{best}}.C_{\text{BB}}$ \KwAnd $C_{\text{curr}}.C_{\text{timing}}$ < $C_{\text{best}}.C_{\text{timing}}$}{
        $\textit{place}_{\textit{best}} \gets\textit{place}_{\textit{curr}}$ \\
      }
    }
  }
  $\alpha \gets$ fraction of attempted moves (out of $N_{\text{moves}}$) that were accepted \\
  $w \gets$ arithmetic average of $\alpha$ across previous five temperature steps \\ 
  $\zeta(w) \gets$ adjust the 3D timing scale using $w$ to control vertical timing emphasis (Section~\ref{sec:meth_var_cost}, Equation~\ref{eq:timing_cost}) \\
  $\theta(w) \gets$ update the timing weight using $w$ to rebalance timing vs. wirelength (Section~\ref{sec:meth_var_cost}, Equation~\ref{eq:tt_equation}) \\
  $T \gets$ decrease the temperature according to the cooling schedule~\cite{vtr_repo} \\
}
\tcp{\textcolor{blue}{Phase 3: Quenching / final refinement}}
\textcolor{black}{$\textit{place}_{\textit{best}} \gets$ perform a final quench (low-temperature refinement) on \textit{$\text{place}_{\text{best}}$}} \\
\Return{\text{$\textit{place}_{\textit{best}}$}}
\end{algorithm}

\noindent
\textbf{Lines 1--14:} The algorithm first partitions the netlist into $k$ layers 
using {\em TritonPart} (Line~2) and constructs an initial 3D placement that respects this layer
assignment (Line~3). Lines~4--10 then initialize all simulated-annealing
parameters. Lines~11--14 evaluate this
initial placement by computing its bounding-box wirelength cost
$C_{\mathrm{BB}}$ and timing cost $C_{\mathrm{timing}}$, and store the
corresponding cost tuple $C_{\text{best}}$ as the current best solution.

\noindent
\textbf{Lines 16--31:} For each temperature level, until the exit temperature is reached (Line~16), exactly
$N_{\text{moves}}$ move attempts are performed (Line~17). Each attempt
samples a 3D-aware move operator from the move set $\mathcal{M}$ using the
move-selection policy of Section~\ref{sec:meth_rl} (Line~18), and computes
the change in total cost $\Delta C_{\text{total}}$ (Lines~18--19).\footnote{We
use the same total-cost formulation as in~\cite{vtr_repo}, except that the
static timing weight $t_f$ is replaced by the dynamic scaling factor
$\theta(w)$, where $w$ is the average move-acceptance
rate from \textcolor{black}{the} previous five temperature steps 
(see Section~\ref{sec:meth_var_cost}).}  The move is accepted if it
improves the total cost or satisfies the Metropolis criterion~\cite{MetropolisRRTT53} 
at the current temperature (Lines~20--21). When a move is accepted, it is applied to the current
placement; the updated $C_{\mathrm{BB}}$ and $C_{\mathrm{timing}}$ are
recomputed; and if the new cost tuple improves upon $C_{\text{best}}$ in both
dimensions, $\texttt{place}_{\text{best}}$ is updated (Lines~22--26).
After all \textcolor{black}{move attempts have been made} at a given temperature \textcolor{black}{level}, $\alpha$ is set to the fraction of
moves that were accepted (Line~27), $w$ is updated as the arithmetic average of
$\alpha$ over the five most recent temperature steps (Line~28), and $\zeta(w)$ and
$\theta(w)$ are adapted based on $w$ to rebalance timing versus wirelength and
to control the emphasis on inter-layer timing effects (Lines~29--30).

\noindent
\textbf{Lines \textcolor{black}{33--34}:} After the annealing process completes, a final low-temperature refinement
(quench) is applied to the best placement to remove residual suboptimal moves. The resulting
placement is returned as the final 3D solution.

\noindent
\textbf{Partition-based layer assignment. }
In 3D FPGA placement, the assignment of logic to layers strongly affects both timing and routability.
A poor initial assignment can force the annealer to spend early iterations repairing poor
cross-layer placements, limiting its ability to explore better 3D configurations later. To avoid
this, we initialize the placement using a structured layer assignment that reflects logical connectivity
and timing criticality. We obtain this assignment by partitioning the packed netlist produced by VTR’s
packing stage~\cite{ElgammalMSMKT25}.\footnote{We partition the packed netlist, rather than the
primitive netlist, because (i) resource usage per layer becomes explicit after packing, and (ii) timing
criticalities are more accurate post-packing.} We build a hypergraph where each vertex represents a
placeable logic block (CLBs, DSPs, BRAMs, I/Os) and each hyperedge represents a net; hyperedges are
weighted using timing criticalities from the packer so that strongly connected or timing-critical
groups tend to remain within the same layer. We then apply \emph{TritonPart} to partition this weighted
hypergraph into $k$ layers ($k=2$ in this work), using a 5\% imbalance factor~\cite{BustanyGKKPW23}.
Finally, blocks are assigned to legal grid locations in \textcolor{black}{decreasing-criticality} order; if a layer becomes
temporarily over-utilized, the block’s layer assignment is relaxed and it is placed on the alternate
layer, ensuring feasibility while preserving the partition-guided structure.

\noindent
\textbf{Placement delay model. }
Timing-driven placement and routing in VTR rely on a delay lookahead model: a precomputed table that
estimates the minimum routing delay required to traverse a given Manhattan offset on the FPGA grid~\cite{MurrayPZWEL20}.
This model guides both the router’s A* search~\cite{HartNR68} and the placer’s timing evaluation, so its accuracy
directly affects final timing quality. In 2D, VTR builds a delay table $delay[\Delta x][\Delta y]$
using Dijkstra’s shortest-path search from a tile\footnote{In VTR, a \emph{tile} is a physical grid
location that may contain a logic block, switch box \textcolor{black}{(SB)}, or connection box \textcolor{black}{(CB)}; it is the fundamental unit
used for placement and routing.} near the bottom-left corner, recording for every reachable tile the
minimum delay observed for that offset~\cite{MurrayPZWEL20}.

For 3D fabrics, this structure extends to a four-dimensional lookup table
$delay[l_{\text{src}}][l_{\text{dst}}][\Delta x][\Delta y]$. However, VTR constructs this table using
a single \emph{average} delay per routing-segment type, causing intra-layer wires, vertical wires, and
inter-layer connections (e.g., TSVs or monolithic \textcolor{black}{(inter-layer) vias (MIVs))} to be treated uniformly—even though TSVs can
be 3–20$\times$ slower and monolithic vias much faster than horizontal wires~\cite{HuangLTCSCK}. To
address this, we modify the delay-model construction to record the \emph{exact per-edge delays}
encountered during Dijkstra’s search rather than relying on segment averages. This allows intra-layer
and inter-layer transitions to be evaluated separately, ensures \textcolor{black}{that} vertical hops reflect true TSV/MIV
delays, and yields more consistent timing estimates for multi-layer nets—all with negligible runtime
overhead.

\subsection{Adaptive cost weighting and refinement}
\label{sec:meth_var_cost}

VTR’s placer uses simulated annealing~\cite{vtr_repo}, beginning with broad 
exploration at high temperature and gradually refining the placement as 
temperature decreases. In 3D placement, two challenges arise:
(i) if cross-layer moves are too permissive early on, \textcolor{black}{the annealer can potentially 
disrupt a good initial layer assignment}; and
(ii) if such moves are over-penalized, the search becomes trapped within layers, 
preventing effective 3D exploration.
Additionally, during the initial annealing stages, timing estimates are unreliable 
because inter-layer connectivity and routing demand are still evolving; over-weighting 
timing at this stage can potentially lead to \textcolor{black}{``suboptimal''} moves, while under-weighting it slows convergence.
To balance these effects, our cost function (Equation~\ref{eq:total_cost}) adapts 
continuously with annealing progress. Specifically, we (i) modulate the effective 
timing cost of cross-layer movement based on annealing state, and (ii) dynamically 
adjust the relative emphasis on timing versus wirelength objectives throughout the process.

\noindent\textbf{Inter-layer timing cost.}

We introduce a scaling factor $\zeta(w)$ that 
dynamically adjusts the timing penalty of vertical transitions during annealing. Early in the 
process, a large $\zeta(w)$ discourages excessive vertical movement and preserves the 
partition-guided layer assignment established at initialization. As annealing progresses and the 
placement stabilizes, $\zeta(w)$ gradually decreases, allowing more cross-layer movement.

This adaptive scaling \textcolor{black}{helps to achieve} a smooth transition from partition-preserving to 
fully 3D timing-driven optimization.
We apply $\zeta(w)$ to the \emph{incremental} vertical timing cost $C_{\text{3D}}(b)$ relative to the 
intra-layer baseline cost $C_{\text{2D}}(b)$:
\begin{align}
\label{eq:timing_cost}
C_{\text{timing}}(b)
= C_{\text{2D}}(b) + \zeta(w)\, C_{\text{3D}}(b),
\end{align}
where
\begin{align*}
C_{\text{2D}}(b) &= \text{delay}[l_{\text{src}}][l_{\text{src}}][\Delta x][\Delta y], \\
C_{\text{3D}}(b) &= \text{delay}[l_{\text{src}}][l_{\text{dst}}][\Delta x][\Delta y]
                   - \text{delay}[l_{\text{src}}][l_{\text{src}}][\Delta x][\Delta y].
\end{align*}
The scale $\zeta(w)$ is driven by the averaged acceptance rate $w$
(to avoid jitter\footnote{Using a moving average for the acceptance rate prevents sharp, step-to-step
changes in $\zeta(w)$, which would otherwise cause abrupt shifts in the cost landscape and destabilize
annealing. The smoothed $w$ lets $\zeta(w)$ track the annealer's progress without noisy fluctuations.})
and follows a monotone, nonincreasing schedule:
\begin{align}
\label{eq:alpha_dynamic}
\zeta(w) &=
\zeta_{\max}
+ (\zeta_{\min} - \zeta_{\max})
\!\left(\frac{w_{\max}-w}{w_{\max}-w_{\min}}\right)^{p_{\zeta}},
\ \ p_{\zeta} \in \{1,2\}.
\end{align}
At high acceptance rates (early, high-temperature stages), a large $\zeta(w)$ imposes strong penalties 
on vertical transitions, effectively constraining placement to a pseudo-2D regime. As $w$ decreases, 
$\zeta(w)$ is gradually reduced toward $\zeta_{\text{min}}$, relaxing penalties and enabling cross-layer moves that refine 
timing on critical paths.

\noindent\textbf{Dynamic timing--wirelength modulation.}
While $\zeta(w)$ modulates local inter-layer timing penalties, the balance between
timing and wirelength in the cost function must also evolve during SA.
Early in the
placement, timing estimates are noisy and the layout is still changing significantly, 
so too much weight on timing can cause the optimizer to chase \textcolor{black}{spurious} gradients. 
We adapt VTR's SA cost function~\cite{ElgammalMSMKT25} and 
use a dynamic schedule $\theta(w)$:
\begin{gather}
C_{\text{total}}(b) \;=\;
(1 - \theta(w))\,C_{\text{BB}}(b)
\;+\; \theta(w)\,C_{\text{timing}}(b),\\
\label{eq:tt_equation}
\theta(w) \;=\;
\begin{cases}
\theta_{\max} - w^{p_{\theta}}\big(\theta_{\max}-\theta_{\min}\big), & \text{if } w > 0.15, \\
\theta_{\max}, & \text{otherwise.}
\end{cases}\\
\label{eq:total_cost}
C_{\text{total}} \;=\; \sum_{b \in \mathcal{B}} C_{\text{total}}(b).
\end{gather}

The SA cost function is Equation~\ref{eq:total_cost}. 
At high acceptance rates ($\alpha$) (exploratory phase), $\theta(w)$ is small, prioritizing wirelength. 
As SA progresses and timing stabilizes, $\theta(w)$ increases smoothly toward $\theta_{\max}$,
shifting emphasis to timing-driven optimization. When $w$ falls below $15\%$, we fix
$\theta(w)=\theta_{\max}$, consistent with~\cite{ElgammalMB22} that characterizes this phase as final annealing
where the ``true objective'' should dominate. \textcolor{black}{$\theta(w)$ is driven in the 
same manner as $\zeta(w)$} to avoid jitter and is constrained 
to be non-decreasing during placement to ensure stable convergence.

\subsection{Expanded SA move set}
\label{sec:meth_rl}

The VTR placer uses a {\em reinforcement learning (RL) agent}~\cite{ElgammalMSMKT25} to adaptively select among several move
types—\emph{uniform}, \emph{centroid}, \emph{median}, and \emph{feasible-region}—each relocating a
single logic block according to a heuristic rule. In centroid moves, a block is displaced toward the
geometric center of its connected blocks; median moves target the median position along each axis;
and weighted variants bias these targets by timing criticality to emphasize critical connections.
Each proposed move is evaluated by the simulated-annealing cost function (Equation~\ref{eq:total_cost}), 
and only accepted moves
update the placement, so directional bias in the move generator directly shapes the explored search
space. In VTR, these moves are augmented with a $z$-coordinate: uniform
moves may target any legal location on any layer; centroid and median moves use the centroid or median
of their neighbors’ layer indices; and feasible-region moves remain confined to the source layer of 
the candidate block.
However, centroid and median moves yield very limited vertical exploration, since an imbalanced layer
distribution among neighbors pulls the averaged $z$-coordinate toward the majority layer (e.g., six
neighbors on layer~1 vs.\ three on layer~2 keeps the target near layer~1), making cross-layer
transitions unlikely unless many neighbors move first to the target layer, under the same biased scheme.

To overcome this limitation, we extend the SA move set to explicitly encourage vertical exploration 
and add a dedicated cross-layer perturbation operator. Centroid and median moves now select the 
destination layer \emph{probabilistically} based on the distribution\footnote{For a block $b$ and $b \in \mathcal B$, let 
$N_b^{(i)}$ be the number of neighbors on layer $i$ and
$N_b^{\text{tot}} = \sum_{j=1}^{k} N_b^{(j)}$. We choose layer $i$ with probability
$P(L=i\mid b) = N_b^{(i)} / N_b^{\text{tot}}$.} of a block’s neighbors across 
layers—e.g., if a block has three neighbors on layer~1 and two on layer~2, it moves to layer~1 with 
probability 60\% and to layer~2 with probability 40\%. 

In addition, we introduce 
a \emph{Layer-Swap} move that exchanges the layer assignments of two blocks while preserving their 
$(x,y)$ coordinates. 
Together, these extensions broaden the action space for 3D placement: adaptive cost 
weighting determines when vertical moves become favorable, and the expanded move set \textcolor{black}{enables} 
beneficial cross-layer \textcolor{black}{perturbations to} be realized once the cost landscape permits them.

\section{Experimental Evaluation}
\label{sec:evaluation}

\begin{figure}[t]
    \centering
    \begin{subfigure}{0.48\linewidth}
        \includegraphics[width=\linewidth]{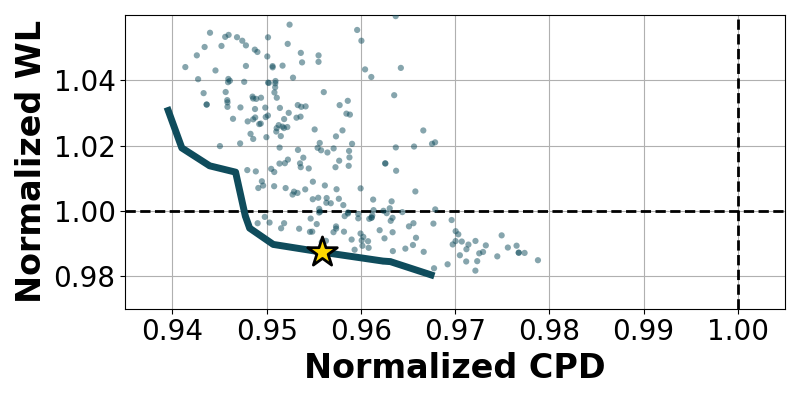}
            \caption{3D CB Pareto Front}
    \end{subfigure}
    \begin{subfigure}{0.48\linewidth}
        \includegraphics[width=\linewidth]{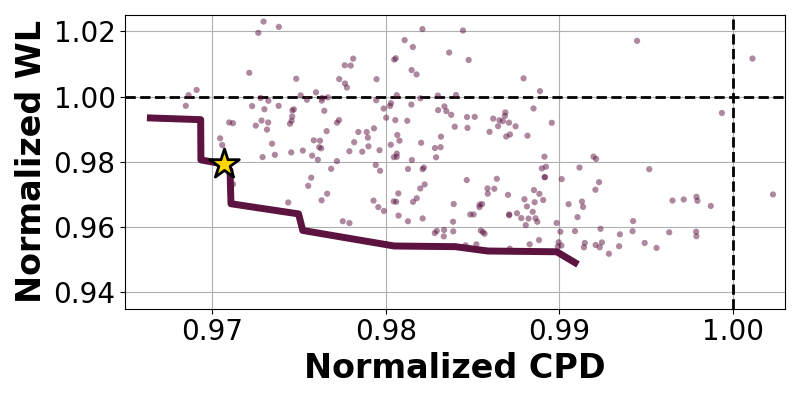}
        \caption{3D CB-O Pareto Front}
    \end{subfigure}
    \vspace{0.5em}
    \begin{subfigure}{0.48\linewidth}
        \includegraphics[width=\linewidth]{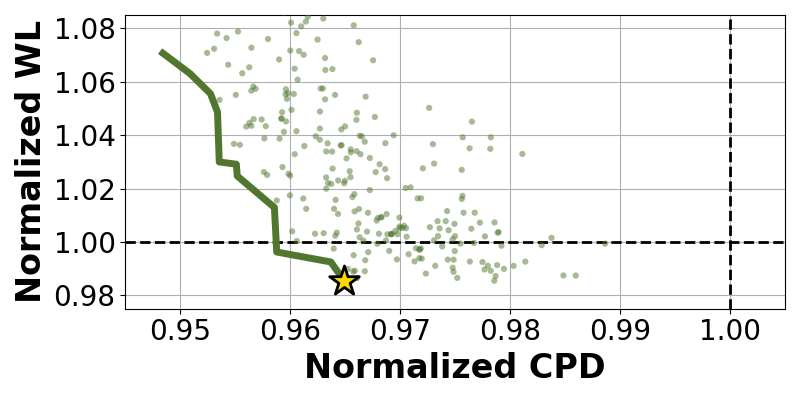}
        \caption{3D CB-I Pareto Front}
    \end{subfigure}
    \begin{subfigure}{0.48\linewidth}
        \includegraphics[width=\linewidth]{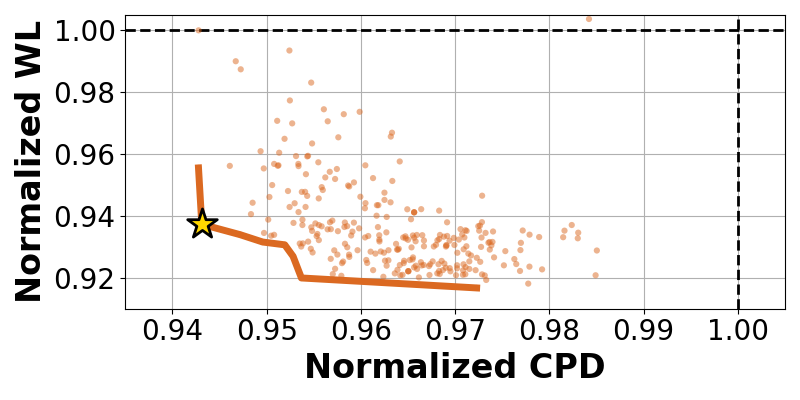}
        \caption{3D SB Pareto Front}
    \end{subfigure}
    \vspace{-8pt}
    \caption{Pareto fronts across the four architectures. CPD and WL are normalized to {\em 3D-baseline}. \textcolor{black}{Stars =  chosen configs.}}
    \label{fig:pareto_fronts}
    \vspace{-8pt}
\end{figure}

We implement our flow in C++ and integrate it into the VTR codebase~\cite{vtr_repo}, using
\emph{TritonPart}~\cite{tritonpart} as the hypergraph partitioner. All experiments are performed on a
server equipped with two 4.1 GHz Intel Xeon Gold 6548Y+ CPUs and 512 GB RAM.
We evaluate our flow using the Koios benchmark suite~\cite{AroraBDMMA23} and the four 3D FPGA architectures based on~\cite{YoussefHH25}.
The FPGA fabric includes I/O blocks along the device periphery, with interior tiles comprising approximately
75\% CLBs, 12.5\% DSPs, and 12.5\% BRAMs. Each benchmark is mapped onto the smallest feasible FPGA grid to ensure high utilization.\footnote{Grid dimensions used are available in the architecture XML files in~\cite{our_flow}.}
Our baselines consist of the VTR implementation in~\cite{vtr_repo} set in two configurations:
(i) \textit{3D-baseline}, VTR run on the above 3D architectures, and
(ii) \textit{2D-baseline}, VTR run on the corresponding \emph{2D-equivalent} architectures.\footnote{A 
2D-equivalent architecture preserves the logical and routing capacity of its 3D counterpart 
while flattening all resources into a single layer.}
Results for critical-path delay (CPD) and wirelength (WL) in all tables and plots are reported relative
to the {\em 2D-baseline} (exceptions are Figures~\ref{fig:pareto_fronts},~\ref{fig:ablation_paretos}) and 
averaged over 10 independent runs with distinct random seeds. For clarity, our
evaluation is structured as follows:
(i) QoR evaluation of our flow (Section~\ref{sec:qor_results});
(ii) hyperparameter exploration (Section~\ref{subsec:hyperparam_explore});
(iii) ablation studies (Section~\ref{subsec:ablation});
(iv) routability evaluation (Section~\ref{tab:min_cw}); and
(v) architecture exploration (Section~\ref{sec:hybrid}).
\textcolor{black}{Full details, data for reproducibility, and \textcolor{black}{an analysis of layer reassignment behavior} are given in~\cite{our_flow}.}

\begin{table*}[t]
\centering
\scriptsize
\caption{QoR comparison: CPD and WL of our flow vs. {\em 3D-baseline}, with all values normalized to {\em 2D-baseline}. 
$\Delta$ is the percentage improvement of our flow compared to the {\em 3D-baseline}. \textcolor{black}{More negative values of $\Delta$ are better.}}
\vspace{-8pt}
\label{tab:koios_full_results}
\setlength{\tabcolsep}{2.5pt}
\renewcommand{\arraystretch}{0.9}
\resizebox{\textwidth}{!}{%
\begin{tabular}{lcccccc cccccc cccccc cccccc}
\toprule
\multirow{3}{*}{\shortstack[c]{\textbf{Benchmark}}} &
\multicolumn{6}{c}{\textbf{3D CB}} &
\multicolumn{6}{c}{\textbf{3D CB-O}} &
\multicolumn{6}{c}{\textbf{3D CB-I}} &
\multicolumn{6}{c}{\textbf{3D SB}} \\
\cmidrule(lr){2-7}
\cmidrule(lr){8-13}
\cmidrule(lr){14-19}
\cmidrule(lr){20-25}
& \multicolumn{3}{c}{CPD} & \multicolumn{3}{c}{WL}
& \multicolumn{3}{c}{CPD} & \multicolumn{3}{c}{WL}
& \multicolumn{3}{c}{CPD} & \multicolumn{3}{c}{WL}
& \multicolumn{3}{c}{CPD} & \multicolumn{3}{c}{WL} \\
\cmidrule(lr){2-4}\cmidrule(lr){5-7}
\cmidrule(lr){8-10}\cmidrule(lr){11-13}
\cmidrule(lr){14-16}\cmidrule(lr){17-19}
\cmidrule(lr){20-22}\cmidrule(lr){23-25}
& {\em 3D-base} & Ours & $\Delta$
& {\em 3D-base} & Ours & $\Delta$
& {\em 3D-base} & Ours & $\Delta$
& {\em 3D-base} & Ours & $\Delta$
& {\em 3D-base} & Ours & $\Delta$
& {\em 3D-base} & Ours & $\Delta$
& {\em 3D-base} & Ours & $\Delta$
& {\em 3D-base} & Ours & $\Delta$ \\
\midrule
clstm\_like.large &
0.952 & 0.917 & \textbf{-3.68\%} &
0.815 & 0.794 & \textbf{-2.58\%} &
0.989 & 0.914 & \textbf{-7.58\%} &
1.030 & 0.890 & \textbf{-13.59\%} &
0.913 & 0.906 & \textbf{-0.77\%} &
0.813 & 0.802 & \textbf{-1.35\%} &
1.112 & 0.944 & \textbf{-15.11\%} &
1.082 & 1.042 & \textbf{-3.70\%} \\
clstm\_like.medium &
0.914 & 0.885 & \textbf{-3.17\%} &
0.823 & 0.811 & \textbf{-1.46\%} &
0.947 & 0.882 & \textbf{-6.86\%} &
0.987 & 0.898 & \textbf{-9.02\%} &
0.899 & 0.881 & \textbf{-2.00\%} &
0.825 & 0.805 & \textbf{-2.42\%} &
1.018 & 0.905 & \textbf{-11.10\%} &
1.086 & 1.055 & \textbf{-2.85\%} \\
clstm\_like.small &
0.941 & 0.948 & +0.74\% &
0.810 & 0.798 & \textbf{-1.48\%} &
0.941 & 0.928 & \textbf{-1.38\%} &
0.935 & 0.895 & \textbf{-4.28\%} &
0.947 & 0.928 & \textbf{-2.01\%} &
0.806 & 0.793 & \textbf{-1.61\%} &
0.992 & 0.993 & +0.10\% &
1.077 & 1.044 & \textbf{-3.06\%} \\
dla\_like.medium &
1.010 & 0.970 & \textbf{-3.96\%} &
0.792 & 0.799 & +0.88\% &
0.968 & 0.970 & +0.21\% &
0.892 & 0.897 & +0.56\% &
1.005 & 0.963 & \textbf{-4.18\%} &
0.788 & 0.805 & +2.16\% &
0.958 & 0.931 & \textbf{-2.82\%} &
1.091 & 1.054 & \textbf{-3.39\%} \\
dla\_like.small &
0.972 & 0.942 & \textbf{-3.09\%} &
0.805 & 0.802 & \textbf{-0.37\%} &
0.953 & 0.938 & \textbf{-1.57\%} &
0.904 & 0.908 & +0.44\% &
0.961 & 0.949 & \textbf{-1.25\%} &
0.797 & 0.805 & +1.00\% &
0.931 & 0.935 & +0.43\% &
1.152 & 1.085 & \textbf{-5.82\%} \\
lstm &
0.907 & 0.910 & +0.33\% &
0.789 & 0.779 & \textbf{-1.27\%} &
0.870 & 0.833 & \textbf{-4.25\%} &
0.964 & 0.953 & \textbf{-1.14\%} &
0.895 & 0.911 & +1.79\% &
0.778 & 0.787 & +1.16\% &
0.931 & 0.809 & \textbf{-13.10\%} &
1.091 & 1.076 & \textbf{-1.37\%} \\
tpu\_like.large.os &
0.893 & 0.842 & \textbf{-5.71\%} &
0.814 & 0.825 & +1.35\% &
0.887 & 0.858 & \textbf{-3.27\%} &
0.861 & 0.902 & +4.76\% &
0.885 & 0.832 & \textbf{-5.99\%} &
0.820 & 0.839 & +2.32\% &
0.827 & 0.797 & \textbf{-3.63\%} &
1.096 & 1.084 & \textbf{-1.09\%} \\
tpu\_like.large.ws &
0.928 & 0.919 & \textbf{-0.97\%} &
0.812 & 0.812 & 0.00\% &
0.903 & 0.888 & \textbf{-1.66\%} &
0.984 & 0.917 & \textbf{-6.81\%} &
0.921 & 0.893 & \textbf{-3.04\%} &
0.801 & 0.805 & +0.50\% &
0.961 & 0.895 & \textbf{-6.87\%} &
1.079 & 1.050 & \textbf{-2.69\%} \\
tpu\_like.small.os &
0.913 & 0.893 & \textbf{-2.19\%} &
0.854 & 0.844 & \textbf{-1.17\%} &
0.887 & 0.856 & \textbf{-3.49\%} &
0.912 & 0.936 & +2.63\% &
0.923 & 0.886 & \textbf{-4.01\%} &
0.836 & 0.846 & +1.20\% &
0.900 & 0.850 & \textbf{-5.56\%} &
1.152 & 1.122 & \textbf{-2.60\%} \\
tpu\_like.small.ws &
1.006 & 0.982 & \textbf{-2.39\%} &
0.805 & 0.794 & \textbf{-1.37\%} &
0.985 & 0.949 & \textbf{-3.65\%} &
0.911 & 0.899 & \textbf{-1.32\%} &
1.001 & 0.973 & \textbf{-2.80\%} &
0.800 & 0.798 & \textbf{-0.25\%} &
1.041 & 0.998 & \textbf{-4.13\%} &
1.073 & 1.053 & \textbf{-1.86\%} \\
bnn &
0.941 & 0.907 & \textbf{-3.61\%} &
0.758 & 0.770 & +1.58\% &
0.934 & 0.913 & \textbf{-2.25\%} &
0.989 & 0.862 & \textbf{-12.84\%} &
0.947 & 0.913 & \textbf{-3.59\%} &
0.765 & 0.763 & \textbf{-0.26\%} &
0.962 & 0.890 & \textbf{-7.48\%} &
0.997 & 0.948 & \textbf{-4.91\%} \\
gemm\_layer &
0.824 & 0.838 & +1.70\% &
0.801 & 0.790 & \textbf{-1.37\%} &
0.862 & 0.837 & \textbf{-2.90\%} &
1.135 & 0.927 & \textbf{-18.33\%} &
0.855 & 0.853 & \textbf{-0.23\%} &
0.786 & 0.788 & +0.25\% &
0.967 & 0.798 & \textbf{-17.48\%} &
1.080 & 1.024 & \textbf{-5.19\%} \\
attention\_layer &
1.050 & 0.973 & \textbf{-7.33\%} &
0.830 & 0.823 & \textbf{-0.84\%} &
0.977 & 0.975 & \textbf{-0.20\%} &
0.882 & 0.882 & 0.00\% &
1.040 & 0.984 & \textbf{-5.38\%} &
0.838 & 0.828 & \textbf{-1.19\%} &
1.026 & 0.964 & \textbf{-6.04\%} &
1.142 & 1.054 & \textbf{-10.16\%} \\
conv\_layer &
1.024 & 0.987 & \textbf{-3.61\%} &
0.820 & 0.817 & \textbf{-0.37\%} &
0.987 & 0.989 & +0.20\% &
0.918 & 0.914 & \textbf{-0.44\%} &
1.024 & 0.986 & \textbf{-3.71\%} &
0.821 & 0.817 & \textbf{-0.49\%} &
1.043 & 1.014 & \textbf{-2.78\%} &
1.208 & 1.099 & \textbf{-9.02\%} \\
spmv &
1.009 & 0.943 & \textbf{-6.54\%} &
0.830 & 0.816 & \textbf{-1.69\%} &
0.962 & 0.926 & \textbf{-3.74\%} &
0.901 & 0.908 & +0.78\% &
0.967 & 0.956 & \textbf{-1.14\%} &
0.824 & 0.818 & \textbf{-0.73\%} &
1.037 & 0.948 & \textbf{-8.58\%} &
1.157 & 1.101 & \textbf{-4.84\%} \\
robot\_rl &
1.017 & 0.979 & \textbf{-3.74\%} &
0.825 & 0.807 & \textbf{-2.18\%} &
0.986 & 0.968 & \textbf{-1.83\%} &
0.966 & 0.954 & \textbf{-1.24\%} &
1.008 & 0.994 & \textbf{-1.39\%} &
0.828 & 0.813 & \textbf{-1.81\%} &
1.004 & 1.004 & 0.00\% &
1.259 & 1.155 & \textbf{-8.26\%} \\
reduction\_layer &
0.878 & 0.853 & \textbf{-2.85\%} &
0.768 & 0.754 & \textbf{-1.82\%} &
0.892 & 0.916 & +2.69\% &
1.291 & 0.880 & \textbf{-31.84\%} &
0.838 & 0.838 & 0.00\% &
0.758 & 0.763 & +0.66\% &
0.876 & 0.863 & \textbf{-1.48\%} &
1.128 & 1.092 & \textbf{-3.19\%} \\
softmax &
1.043 & 0.979 & \textbf{-6.14\%} &
0.816 & 0.823 & +0.86\% &
0.976 & 0.974 & \textbf{-0.20\%} &
0.926 & 0.932 & +0.65\% &
1.066 & 0.976 & \textbf{-8.44\%} &
0.812 & 0.824 & +1.48\% &
1.032 & 0.988 & \textbf{-4.26\%} &
1.173 & 1.068 & \textbf{-8.95\%} \\
conv\_layer\_hls &
1.057 & 1.000 & \textbf{-5.39\%} &
0.819 & 0.809 & \textbf{-1.22\%} &
0.986 & 0.992 & +0.61\% &
0.924 & 0.862 & \textbf{-6.71\%} &
1.049 & 1.016 & \textbf{-3.15\%} &
0.827 & 0.788 & \textbf{-4.72\%} &
1.050 & 0.994 & \textbf{-5.33\%} &
1.088 & 1.015 & \textbf{-6.71\%} \\
eltwise\_layer &
1.051 & 1.002 & \textbf{-4.66\%} &
0.812 & 0.788 & \textbf{-2.96\%} &
0.995 & 0.963 & \textbf{-3.22\%} &
0.903 & 0.910 & +0.78\% &
1.015 & 1.001 & \textbf{-1.38\%} &
0.806 & 0.798 & \textbf{-0.99\%} &
1.028 & 0.958 & \textbf{-6.81\%} &
1.199 & 1.110 & \textbf{-7.42\%} \\
\midrule
\textbf{Geomean} &
0.964 & 0.932 & \textbf{-3.32\%} &
0.810 & 0.803 & \textbf{-0.86\%} &
0.943 & 0.922 & \textbf{-2.23\%} &
0.956 & 0.906 & \textbf{-5.52\%} &
0.956 & 0.930 & \textbf{-2.72\%} &
0.806 & 0.804 & \textbf{-0.25\%} &
0.982 & 0.921 & \textbf{-6.21\%} &
1.119 & 1.066 & \textbf{-4.74\%} \\
\bottomrule
\end{tabular}
}
\end{table*}

\subsection{QoR comparison}
\label{sec:qor_results}
Table~\ref{tab:koios_full_results} presents CPD and WL results compared to the
{\em 3D-} and {\em 2D-} baselines. We run 10 placement trials with
different random seeds and report the arithmetic mean, with all values
normalized to the {\em 2D-baseline}. Across all four 3D architectures, our
flow consistently outperforms both baselines. 

Geometric-mean improvements of our flow compared to the
{\em 3D-}({\em 2D-}) baselines are as follows\textcolor{black}{:}
(i) 3D~CB: CPD improves by 3.32\% (6.80\%) and WL by 0.86\% (19.70\%);
(ii) 3D~CB-O: CPD improves by 2.23\% (7.80\%) and WL by 5.52\% (9.40\%);
(iii) 3D~CB-I: CPD improves by 2.72\% (7.00\%) and WL by 0.25\% (19.60\%); \textcolor{black}{and}
(iv) 3D~SB: CPD improves by 6.21\% (7.90\%) and WL by 4.74\%. 

The 3D SB architecture benefits \textcolor{black}{the} most from our flow because its vertical connections are assigned
dynamically; 
without an accurate 3D delay model, it
is particularly sensitive to lookahead inaccuracies, since the delay lookahead matrix~\cite{vtr_repo}
also guides routing-path exploration (Section~\ref{sec:methodology}).

\begin{table}[t]
\centering
\caption{Optimal \textcolor{black}{parameter} set found for each architecture.}
\vspace{-8pt}
\label{tab:optimal_params}
\scriptsize
\renewcommand{\arraystretch}{0.9}
\begin{tabular}{lcccc}
\toprule
\textbf{Parameter} & \textbf{3D CB} & \textbf{3D CB-O} & \textbf{3D CB-I} & \textbf{3D SB} \\
\midrule
$p_{\zeta}$ & 1 & 1 & 1 & 2 \\
$p_{\theta}$ & 1 & 1 & 1 & 1 \\
$\theta_{min}$ & 0.03 & 0.09 & 0.03 & 0.35 \\
$\theta_{max}$ & 0.51 & 0.80 & 0.51 & 0.79 \\
$\zeta_{max}$ & 1.6 & 1.4 & 2.8 & 2.0 \\
$\zeta_{min}$ & 1.0 & 1.0 & 1.0 & 1.0 \\
$w_{max}$ & 0.41 & 0.31 & 0.79 & 0.26 \\
$w_{min}$ & 0.32 & 0.16 & 0.61 & 0.15 \\
\bottomrule
\end{tabular}

\end{table}

\subsection{Hyperparameter exploration}
\label{subsec:hyperparam_explore}
We tune \textcolor{black}{the} hyperparameters controlling the adaptive cost functions $\zeta(w)$ and $\theta(w)$ 
(Section~\ref{sec:meth_var_cost}) using \texttt{Optuna}'s Tree-structured Parzen Estimator (TPE) 
sampler~\cite{Watanabe23}, and jointly optimize WL and CPD. Each trial evaluates 
one configuration on nine representative Koios designs across five random seeds, 
\textcolor{black}{and we perform} 300 trials per architecture; for each seed, CPD and WL are measured, normalized to the 
{\em 3D-baseline}, and aggregated via geometric mean as the trial objective. 
\textcolor{black}{The resulting 
Pareto fronts are shown in Figure~\ref{fig:pareto_fronts}.} 
\textcolor{black}{For each architecture, 
we choose the Pareto-efficient configuration that 
maximizes combined gain in CPD and WL (yellow star in each plot). 
The selected configurations 
are summarized in Table~\ref{tab:optimal_params}.}

\begin{figure}[t]
    \centering
    \includegraphics[width=0.9\columnwidth]{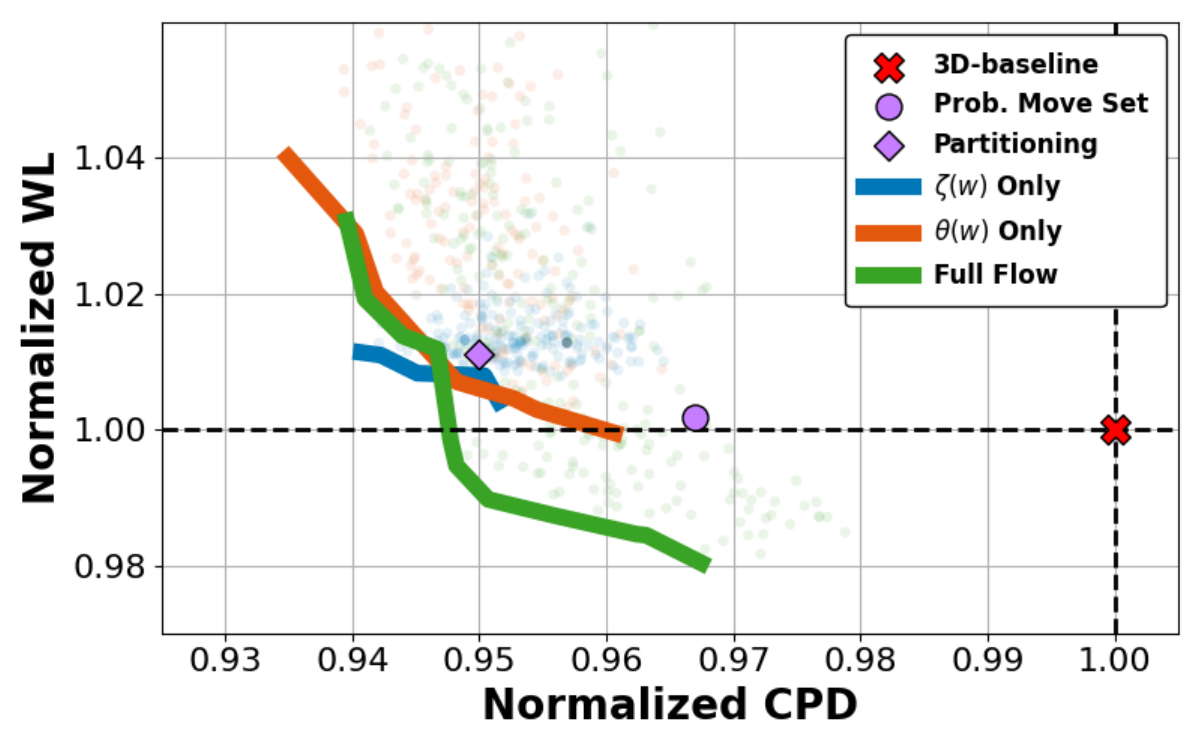}
    \vspace{-10pt}
    \caption{Ablation studies on \textcolor{black}{a} subset of the Koios benchmarks.}
    \label{fig:ablation_paretos}
    \vspace{-8pt}
\end{figure}

\subsection{Ablation studies}
\label{subsec:ablation}

To quantify the contribution of each component in our 3D placement flow, we perform ablation studies 
on the CB architecture using the same Optuna-based hyperparameter tuning setup described in 
Section~\ref{subsec:hyperparam_explore}, but enabling only one enhancement at a time. {\em Partition-based 
layer assignment} and the {\em probabilistic move-set} expansion are ``binary features'' and therefore appear 
in Figure~\ref{fig:ablation_paretos} as individual points, whereas the {\em 3D timing-scaling factor} $\zeta(w)$, 
the {\em timing–wirelength balance} $\theta(w)$, and the {\em full flow} produce Pareto curves as their hyperparameters vary. 
Results show that tuning $\zeta(w)$ (blue) and $\theta(w)$ (orange) can achieve CPD improvements 
comparable to those of the {\em full flow}, but only the {\em full flow} simultaneously improves both CPD and WL 
relative to the {\em 3D-baseline}. This indicates that the \textcolor{black}{{\em combination}} 
of \textcolor{black}{our} proposed enhancements 
is \textcolor{black}{needed} to obtain consistent timing and wirelength gains.

\begin{figure}[bt]
    \centering
    \includegraphics[width=\columnwidth]{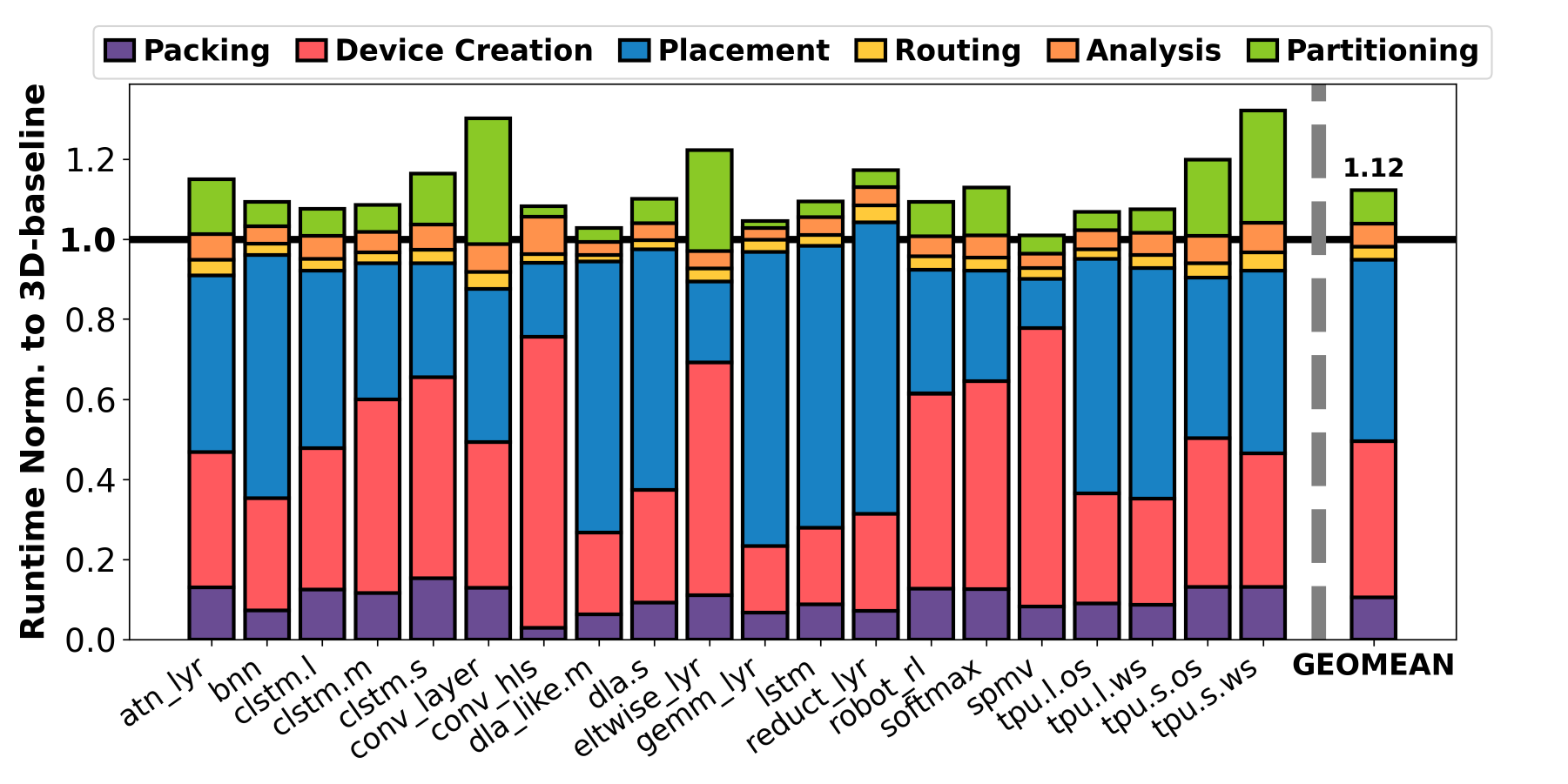}
    \vspace{-16pt}
    \caption{Runtime profiling on 3D CB. Results are normalized to {\em 3D-baseline} and averaged over 10 seeds. }
    \label{fig:runtime_result}
    \vspace{-8pt}
\end{figure}

\noindent
\textbf{Runtime remarks.}
Figure~\ref{fig:runtime_result} illustrates the runtime overhead of our flow on the Koios benchmarks, 
normalized to {\em 3D-baseline}. Overall, our flow incurs a 12\% increase in 
total runtime (geometric mean). \textcolor{black}{As the stacked breakdown shows, 
the runtime increases above the} 
baseline (i.e., exceeding the $y=1$ line) is dominated by the partitioning stage, 
which accounts for most of the additional runtime overhead.

\begin{table}[t]
\centering
\caption{Comparison of \textit{min\_CW} of our flow vs. {\em 3D-baseline}. Architecture is 3D CB-O; results \textcolor{black}{are} averaged over 10 seeds.}
\vspace{-8pt}
\label{tab:min_cw}
\tiny

\renewcommand{\arraystretch}{0.9}
\begin{tabular}{lccclccc}
\toprule
\textbf{Benchmark} & \textbf{VTR 9} & \textbf{Ours} & \textbf{$\Delta$} &
\quad
\textbf{Benchmark} & \textbf{VTR 9} & \textbf{Ours} & \textbf{$\Delta$} \\
\midrule

attention\_layer      & 153.6 & 154.8 & +0.78\% 
& bnn                   & 95.0  & 115.8 & +21.89\% \\

clstm\_like.large     & 109.0 & 119.6 & +9.72\%  
& clstm\_like.medium    & 107.6 & 102.6 & \textbf{-4.65\%} \\

clstm\_like.small     & 111.6 & 110.8 & \textbf{-0.72\%} 
& conv\_layer           & 199.6 & 189.8 & \textbf{-4.91\%} \\

conv\_layer\_hls      & 67.2  & 66.6  & \textbf{-0.89\%}
& dla\_like.medium      & 201.8 & 224.2 & +11.10\% \\

dla\_like.small       & 182.4 & 171.0 & \textbf{-6.25\%}
& eltwise\_layer        & 101.4 & 94.2  & \textbf{-7.10\%} \\

gemm\_layer           & 100.4 & 96.6  & \textbf{-3.78\%}
& lstm                  & 155.4 & 150.6 & \textbf{-3.09\%} \\

reduction\_layer      & 90.4  & 62.8  & \textbf{-30.53\%}
& robot\_rl             & 120.2 & 102.4 & \textbf{-14.81\%} \\

softmax               & 72.8  & 76.6  & +5.22\%
& spmv                 & 89.2  & 87.6  & \textbf{-1.79\%} \\

tpu\_like.large.os    & 148.8 & 135.6 & \textbf{-8.87\%}
& tpu\_like.large.ws    & 154.8 & 162.8 & +5.17\% \\

tpu\_like.small.os    & 130.4 & 129.2 & \textbf{-0.92\%}
& tpu\_like.small.ws    & 155.3 & 139.2 & \textbf{-10.37\%} \\

\midrule
\scriptsize{\textbf{Geomean}} & & &  
& & & & \scriptsize{\textbf{-2.82\%}}\\
\bottomrule
\end{tabular}

\end{table}

\subsection{Routability evaluation}

Routability is a critical measure of placement quality, since poor spatial organization can lead to excessive
routing demand or even unroutable designs. We assess routability by comparing the minimum channel width\footnote{Minimum 
channel width (\textit{min\_CW}) 
is the smallest number of routing tracks per routing {\em channel} 
that allows all nets to be routed; lower \textit{min\_CW} indicates better routability and 
more balanced congestion~\cite{BetzRM99}.} 
(\textit{min\_CW}) required for successful routing between our 3D placement flow and the {\em 3D-baseline}. As summarized in
Table~\ref{tab:min_cw} for the Koios benchmarks on the 3D~CB-O architecture (averaged over ten seeds), our flow
reduces \textit{min\_CW} by 2.82\% on average (geo-mean) relative to {\em 3D-baseline}, with the \texttt{reduction\_layer}
showing the most (30.53\%) reduction.

\subsection{Architecture exploration}
\label{sec:hybrid}

\begin{figure}[t]
    \centering
    \includegraphics[width=1.0\columnwidth]{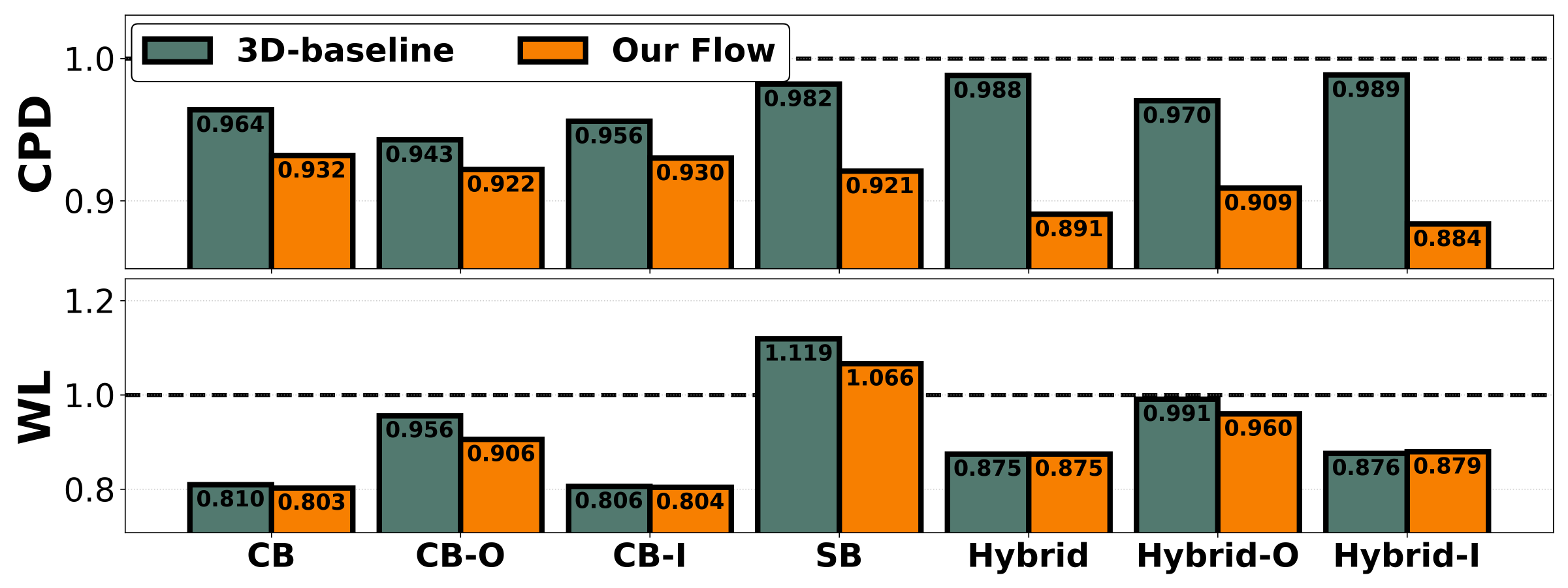}
    \vspace{-20pt}
    \caption{Results of various hybrid architectures on \textcolor{black}{{\em 3D-baseline}} 
    and our flow, normalized to \textcolor{black}{{\em 2D-baseline}}.}
    \label{fig:hybrid_results}
    \vspace{-12pt}
\end{figure}

To assess the generality of our flow and its utility for architectural studies, we evaluate a set of
\emph{hybrid architectures} that combine vertical connectivity through both logic pins and switch-box
routing. Unlike the base CB, CB-O, CB-I, and SB architectures—where vertical links are restricted to a
single connection type—hybrid architectures enable richer 3D connectivity (illustrated in
Figure~\ref{fig:archs}). We study three variants:
(i) \textit{Hybrid}, enabling both logic-pin and \textcolor{black}{SB} vertical connections;
(ii) \textit{Hybrid-O}, enabling output-pin and \textcolor{black}{SB} connections; and
(iii) \textit{Hybrid-I}, enabling input-pin and \textcolor{black}{SB} connections.

\begin{figure}[bt]
    \centering
    \includegraphics[width=\columnwidth]{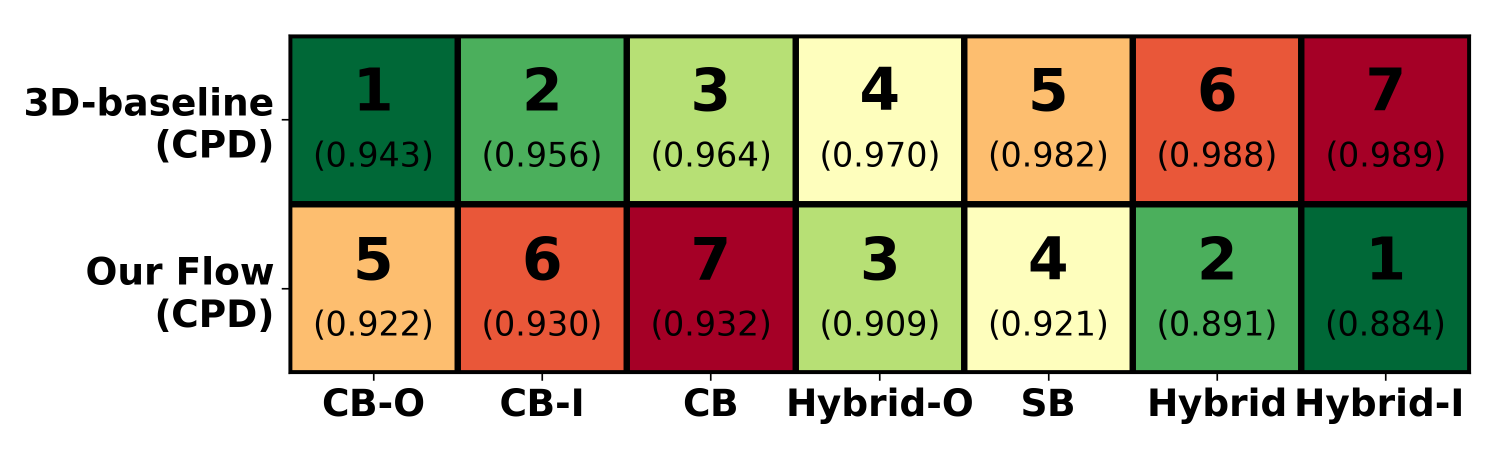}
    \vspace{-20pt}
    \caption{Relative CPD ordering of different 3D architectures on Koios benchmarks using 
    \textcolor{black}{{\em 3D-baseline}} and our flow.}
    \label{fig:ordering_heatmap}
    \vspace{-20pt}
\end{figure}

Figure~\ref{fig:hybrid_results} \textcolor{black}{shows} normalized QoR (geometric mean across all Koios designs)
using both {\em 3D-baseline} and our flow. With {\em 3D-baseline}, hybrid architectures appear inferior to the base 3D
architectures, reflecting the placer’s limited ability to exploit vertical flexibility. In contrast,
our 3D-aware flow unlocks \textcolor{black}{the potential of the hybrid
architectures}: CPD improves by up to 10.6\%
(Hybrid-I) and WL by up to 3.13\% (Hybrid-O) relative to {\em 3D-baseline}.
Figure~\ref{fig:ordering_heatmap} further highlights this shift in architectural ranking. With
{\em 3D-baseline}, CB-O appears to be the strongest-performing architecture. With our flow, the hybrid
architectures rise to the top, outperforming all base architectures. These results indicate that
previous assessments understated the benefits of hybrid connectivity, and demonstrate that our
placement framework provides the accuracy needed for meaningful and ``directionally-correct''
3D FPGA architecture exploration.

\section{Conclusion}
\label{sec:conclusion}
We \textcolor{black}{have} presented an open-source 3D FPGA placement flow that 
combines partitioning-based initialization, refined
3D delay modeling, adaptive cost scheduling, and expanded 3D move sets. These enhancements improve timing
guidance and enable \textcolor{black}{effective} exploration of the vertical design space. Across multiple 3D architectures,
our flow achieves up to \textbf{10.6\%} lower critical-path delay and \textbf{5.52\%} shorter wirelength
relative to \textcolor{black}{VTR~9}, while preserving routability.
\textcolor{black}{Ongoing efforts include extending the flow to multi-layer ($>2$) 
architectures, incorporating thermal
and power models, and exploring 3D routing optimizations. Our work thus
provides foundations for the advance of 3D FPGA CAD and architecture research.}

\begin{acks}
Research at UCSD is supported in part by Altera. Research at Georgia Tech was partially supported by the NSF under Grant Numbers 2338365 and 2317251.
\end{acks}


\def\baselinestretch{}

\end{document}